\documentclass[prb,aps,twocolumn,superscriptaddress]{revtex4-2}

\usepackage{graphicx,color}
\usepackage{amsthm}
\usepackage{amsfonts}
\usepackage{algorithmic}
\usepackage{enumerate}
\usepackage{latexsym}
\usepackage{amsmath}
\usepackage{amssymb}
\usepackage{ulem}
\usepackage{color}
\usepackage[colorlinks=true,citecolor=blue,urlcolor=blue,linkcolor=blue]{hyperref}

\emergencystretch=\maxdimen
\include{MyCommand}

\begin{document}
\title{Disorder Suppression of Charge Density Waves in the Honeycomb Holstein Model}
\author{Guangchao Li}
\affiliation{School of Physical Science and Technology, Beijing University of Posts and Telecommunications, Beijing 100876, China}
\author{Lifei Zhang}
\affiliation{School of Physical Science and Technology, Beijing
University of Posts and Telecommunications, Beijing 100876, China}
\author{Tianxing Ma}
\affiliation{School of Physics and Astronomy, Beijing Normal University, Beijing 100875, China}
\affiliation{Key Laboratory of Multiscale Spin Physics (Ministry of Education), Beijing Normal University, Beijing 100875, China}
\author{Qionglin Dai}
\affiliation{School of Physical Science and Technology, Beijing
University of Posts and Telecommunications, Beijing 100876, China}
\author{Lufeng Zhang}
\email{lfzhang@bupt.edu.cn}
\affiliation{School of Physical Science and Technology, Beijing University of Posts and Telecommunications, Beijing 100876, China}

\begin{abstract}
The formation of charge-density-wave order in Dirac fermion systems via electron-phonon coupling represents a significant topic in condensed matter physics. In this work, we investigate this phenomenon within the Holstein model on the honeycomb lattice, with a specific focus on the effect of disorder. While the interplay between electron-electron interactions and disorder has long been a central theme in the field, recent attention has increasingly turned to the combined influence of disorder and electron-phonon coupling. Using determinant quantum Monte Carlo simulations, we concentrate on the phase transitions of charge-density-wave order on the honeycomb lattice. Disorder is introduced through the random hopping of electrons in the system, which can localize electrons via the Anderson effect. Our primary result is that disorder suppresses the charge-density-wave phase, and the  interplay between disorder and electron-phonon interactions extends the phase area. We also determine the transition temperature \(\beta_c\) to the ordered phase as a function of the electron-phonon coupling. Additionally, we observed a suppression of electron kinetic energy and dc conductivity under disorder, highlighting the role of Anderson localization in the degradation of electronic transport. These findings offer significant theoretical insight into the stability and critical phenomena of correlated phases in disordered two-dimensional systems.
\end{abstract}

\pacs{71.10.Fd, 74.20.Rp, 74.70.Xa, 75.40.Mg}

\maketitle


\section{Introduction}
The isolation and manipulation of monolayer graphene have elevated Dirac fermions to a central position in contemporary condensed matter theory \cite{doi:10.1126/science.1102896}. Although single-particle properties in such systems are well established within the framework of relativistic band structures, the role of electron correlations remains both subtle and profoundly rich, motivating ongoing theoretical exploration \cite{RevModPhys.81.109,RevModPhys.82.2673,RevModPhys.83.407,RevModPhys.84.1067,Geim2007,Black-Schaffer_2014,PhysRevLett.106.236805}. The two-dimensional (2D) nature of these systems improves the efficiency of a wide range of analytical and numerical techniques ranging from renormalization group analyses to large-scale quantum Monte Carlo simulations allowing systematic studies of interaction-driven quantum criticality and emergent collective phases \cite{PhysRevB.110.235128,PhysRevB.110.085103,PhysRevB.109.045107,PhysRevB.106.235140,PhysRevB.106.205149,PhysRevLett.120.116601,PhysRevB.110.L220506}. Recent discoveries in related honeycomb lattice materials have further invigorated the field: quantum spin Hall insulators in bismuthene exemplify the interplay of topology and spin-orbit coupling \cite{PhysRevB.104.205105,PhysRevB.94.014115}, while twisted bilayer graphene exhibits correlated insulating states and unconventional superconductivity near magic angles \cite{PhysRevB.103.115431,PhysRevB.98.241407,PhysRevB.110.115154}, challenging conventional paradigms of pairing mechanisms. Moreover, massive Dirac systems such as charge-density-wave (CDW) phases in transition-metal dichalcogenides illustrate how spontaneous symmetry breaking can gap out Dirac spectra \cite{PhysRevB.108.224103,PhysRevLett.132.226401}, offering a platform to investigate interaction-induced mass generation and its potential applications in optoelectronics and quantum devices. While the competition between superconductivity and charge order is well-documented in clean systems with various electron-phonon couplings \cite{PhysRevB.97.140501,PhysRevLett.127.247203,rhss-d52m,PhysRevLett.134.206001}, the fate of this delicate balance in the presence of strong disorder—a key feature of real materials remains a pivotal open question, which we address in this work.

The interplay between disorder and electronic correlations has long been a cornerstone of condensed matter physics, epitomized by Anderson localization and its profound impact on electron transport and phase stability \cite{PhysRevLett.42.673,Wegner1980,RevModPhys.57.287,PhysRevB.37.325,RevModPhys.66.261,doi:10.1126/science.1107559,RevModPhys.78.373,Vojta2019Disorder,PhysRevB.103.L060501,S_V_Kravchenko_2004}. In real graphene-based systems, structural imperfections, substrate effects, and intentional impurities introduce disorder that can significantly alter electronic behavior \cite{PhysRevB.66.073102,PhysRevB.85.165117,PhysRevB.89.155434}. When combined with a strong electron-phonon coupling (EPC), such a disorder can destabilize or modulate emergent orders such as CDWs, which rely on delicate electron-lattice coherence \cite{RevModPhys.60.1129,PhysRevB.110.205145,PhysRevLett.122.077602,PhysRevLett.122.077601,PhysRevB.98.085405,PhysRevB.102.161108,PhysRevLett.126.107205}. Despite its importance, a systematic understanding of how disorder influences CDW formation in the presence of EPC especially on a Dirac-fermion system like graphene remains an open question. This work aims to bridge this gap by exploring the suppression and critical behavior of CDW order in the disordered Holstein model on the honeycomb lattice using large-scale determinant quantum Monte Carlo (DQMC) simulations \cite{PhysRevD.24.2278,CREUTZ1981427,PhysRevB.40.197,PhysRevLett.66.778,santos_2003,PhysRevB.41.9301,PhysRevLett.94.170201,PhysRevB.92.045110}.
In Sec.II, we detail the Hamiltonian of the disordered Holstein model, explicitly incorporating electron-phonon interaction terms and random hopping kinetic energy. We outline the DQMC framework, emphasizing its capability to access finite-temperature properties of interacting fermions without the fermion sign problem. In Sec.III, we present numerical insights into how disorder modulates CDW correlations, revealing suppression of long-range order and a shift in critical temperature governed by the interplay between electron-phonon coupling and randomness. Finally, in Sec.IV, we summarize our conclusions, discussing their implications for electron behavior in disordered 2D materials.

\section{Model and method}
The Holstein model provides a theoretical framework for understanding the dynamics of electrons interacting with a local phonon field within a lattice structure. This model is valuable in elucidating the complex relationship between electronic behavior and phonons. It describes the movement of electrons and their interaction with local phonons by means of a site-dependent density, and the Hamiltonian is as follows:
\begin{align}
    \hat{\mathcal{H}} = &-\sum_{\langle \boldsymbol{i,j}\rangle, \sigma} {t}_{\boldsymbol{ij}}(\hat{d}_{\boldsymbol{i}\sigma}^\dagger \hat{d}_{\boldsymbol{j}\sigma} + \text{H.c.})-\mu \sum_{\boldsymbol{i},\sigma} \hat{n}_{\boldsymbol{i},\sigma} \notag \\
    & + \frac{1}{2} \sum_{\boldsymbol{i}} \hat{P}_{\boldsymbol{i}}^2 + \frac{\omega_0^2}{2} \sum_{\boldsymbol{i}} \hat{X}_{\boldsymbol{i}}^2 + \lambda \sum_{\boldsymbol{i},\sigma} \hat{n}_{\boldsymbol{i},\sigma} \hat{X}_{\boldsymbol{i}}
    \label{eu:01}
\end{align}

The first term represents the kinetic energy associated with the electron hopping between nearest-neighbor sites on a two-dimensional lattice. Here, \(t_{\boldsymbol{ij}}\) stands for the hopping integral between sites \(\boldsymbol{i}\) and \(\boldsymbol{j}\). The operators \(\hat{d}_{\boldsymbol{i}\sigma}^{\dagger}\) and \(\hat{d}_{\boldsymbol{j}\sigma}\) are used to create and annihilate electrons with spin \(\sigma\) at sites \(\boldsymbol{i}\) and \(\boldsymbol{j}\), respectively, and these two operators play a crucial role in describing the electron hopping process in the lattice and the related changes in the quantum state. The second term involves the chemical potential \(\mu\), whose main function is to regulate the average number of electrons in the system. \(\hat{n}_{\mathbf{i}\sigma}=\hat{d}_{\mathbf{i}\sigma}^{+}\hat{d}_{\mathbf{i}\sigma}\) represents the number of electrons with spin \(\sigma\) on site \(\mathbf{i}\). The third and fourth terms are related to the phonon energy. Among them, \(\hat{P}_{\boldsymbol{i}}\) and \(\hat{X}_{\boldsymbol{i}}\) are the momentum operator and the displacement operator of the phonon, respectively, and \(\omega_{0}\) is the frequency of the local phonon mode. These quantities jointly determine the energy state of the phonon in the system and its contribution to the overall system energy. The last term reflects the electron-phonon coupling. The coupling constant \(\lambda\) characterizes the intensity of this interaction. It reflects the degree of close mutual influence between electrons and phonons as well as the correlation in aspects such as energy transfer.

Disorder is introduced into the model through the hopping matrix element \(t_{\boldsymbol{ij}}\). \(t_{\boldsymbol{ij}}\) is uniformly selected within the interval \([t-\Delta/2, t+\Delta/2]\). Moreover, the average value of \(t_{\boldsymbol{ij}}\), denoted as \(\overline{t_{\boldsymbol{ij}}}\), is set to \(1\), and its probability distribution \(P(t_{\boldsymbol{ij}})\) is \(1 / \Delta\) within this interval and zero outside this interval. The parameter \(\Delta\) is used to quantify the degree of disorder, and it directly determines the influence of disorder factors in the model.

All the data presented in this paper are simulated by the DQMC method. In the Holstein model, there is no electron-electron interaction, so the fermionic operators in the Hamiltonian are only quadratic, allowing them to be traced out in the partition function expression.  As for the phonon operators, we can convert them within the model into numerical quantities through path integration, thus the original electron-phonon interaction problem transformed into one that involves only fermionic operators. Therefore, in contrast to the Hubbard model on the honeycomb lattice, where the sign problem \cite{PhysRevB.41.9301} emerges when away from half-filling, there is no sign problem in our case. Consequently, low-temperature properties become accessible through DQMC simulations.

To begin with, we study the electronic correlation properties of the system to characterize the CDW phase. The CDW order is defined by calculating the density-density correlation function \(c(r)\) and its Fourier transform, the charge structure factor \(S_{c}(\mathbf{k})\):
\begin{align}
 c(r) = \langle (n_{i\uparrow }+ n_{i\downarrow})(n_{i+r\uparrow }+ n_{i+r\downarrow}) \rangle
 \label{equ:02}
\end{align}
\begin{align}
 S_{c}(\mathbf{k})=\frac{1}{N} \sum_{\mathbf{r}} c(\mathbf{r}) e^{i \mathbf{k} \cdot \mathbf{r}}.
 \label{equ:03}
\end{align}
We predominantly investigate the staggered pattern of charge ordering \(S_{CDW}=\sum_r (-1)^r c(r)\). By analyzing the variation of \(S_{CDW}\) with temperature and electron-phonon coupling strength, we can determine the critical points and phase diagrams of the CDW phase.

Moreover, we also conduct an in-depth investigation into the direct-current conductivity \(\sigma_{dc}\) of the system by computing the current-current correlation function:
\begin{align}
 \sigma_{dc} \approx \frac{\beta^2}{\pi} \chi_{xx}(q=0, \tau = \beta/2)
 \label{equ:04}
\end{align}
Here, \(\chi_{xx}(q, \tau)=\langle j_x(q, \tau) j_x(-q, 0) \rangle\) represents the current-current correlation function, and \(j_x(q, \tau)\) is the Fourier transform of the current operator. This computational approach enables us to gain a profound understanding of the transport properties, particularly under diverse temperature and disorder conditions.

In order to investigate the lattice size effect, the data are presented for $L = 4,5,6$, where \(L\) represents the lattice size, and for the honeycomb lattice, the number of lattice points is \(N = 2L^{2}\). A dimensionless electron-phonon coupling strength \(\lambda_{D}=\lambda^{2}/(\omega_{0}^{2}W)\) is employed, where \(W = 6t\) denotes the bandwidth. To ensure that the systematic Trotter error of the system is smaller than the statistical error of Monte Carlo sampling, the discretization mesh is set as \(\Delta\tau = 1/20\). In the Holstein model, \(\mu =-\lambda^{2}/\omega_{0}^{2}\) corresponds to the half-filling case. We set the parameter \(\mu\) to maintain the charge density \(\rho\) in a half-filled state \(\rho=\langle \hat{n}_{\mathbf{i}} \rangle = 1\). We treat disorder as quenched (static). Independent static realizations of the random hopping configuration are generated. This models a frozen disorder potential and is the standard approach for studying static disorder effects, in contrast to treating disorder as a dynamical (annealed) field. For each realization, observables are computed across all temperatures using the same fixed configuration, isolating the temperature dependence. All data points are the average of 20 independent disorder configurations; the error bars correspond to the standard deviation of these 20 datasets, which comprehensively reflects both Monte Carlo sampling fluctuations and disorder configurational fluctuations\cite{PhysRevLett.120.116601}.

\section{Results and discussion}
\subsection{The charge-density correlation function}
\begin{figure}[t]
\centering
\includegraphics[width=1.0\columnwidth]{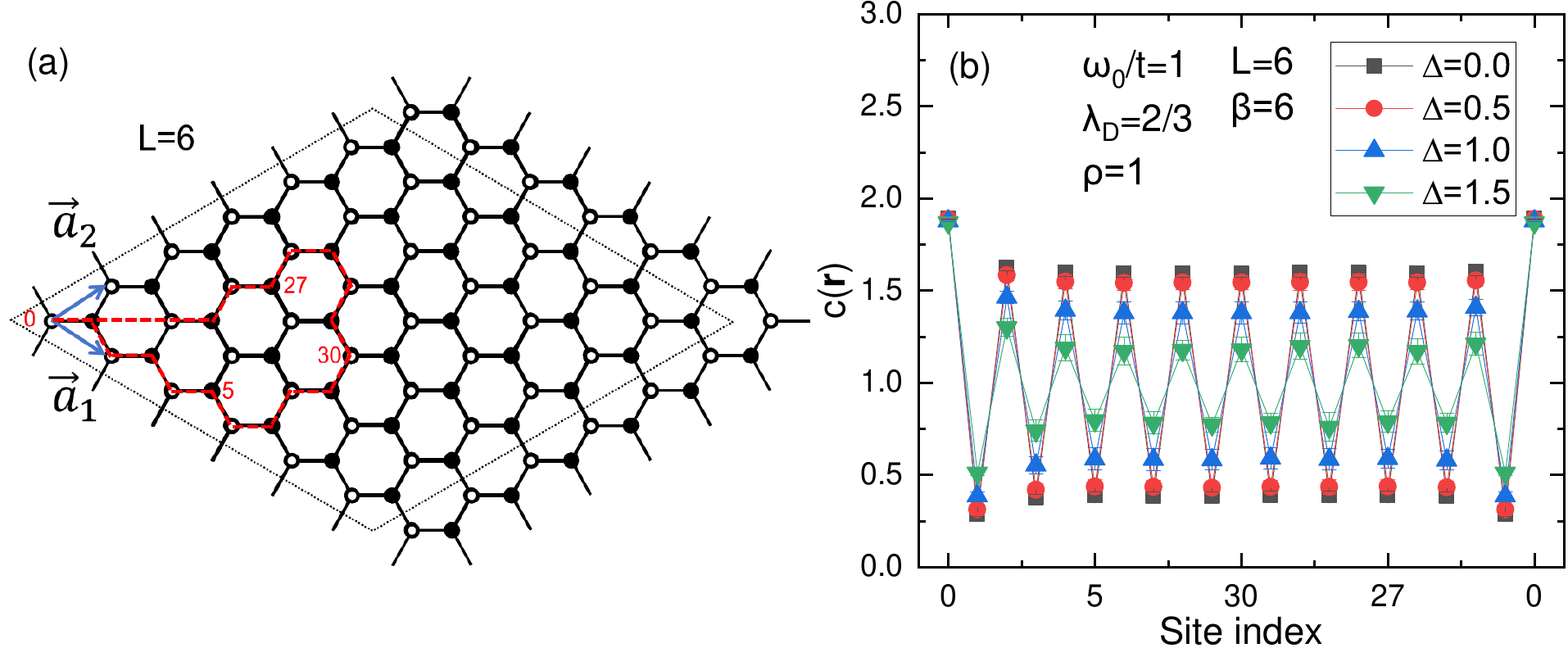}
\caption{(a) A honeycomb lattice structure with \(L = 6\). \(\vec{a}_{1}\) and \(\vec{a}_{2}\) are the two primitive vectors of the lattice, which define the periodic structure of the honeycomb lattice. The black solid dots and white hollow dots represent the two sub-lattices of the honeycomb lattice. The trajectory (the red dashed line) corresponds to the horizontal axis of plot (b). (b) The influence of different disorder strengths \(\Delta\) (taking \(\Delta =0.0\), \(\Delta =0.5\), \(\Delta =1.0\), \(\Delta =1.5\) respectively, distinguished by markers of different colors and shapes) on the charge-correlation function \(c(\boldsymbol{r})\). Under the conditions of a system size \(L = 6\) and fixed parameters \(\omega_{0}/t=1\), \(\lambda_{D}=2/3\), \(\beta=6\), \(\mu=-4\).} %
 \label{fig:01}
 \end{figure}

We initially present the density-density correlation function \(c(\boldsymbol{r})\) along the real-space path in Fig. \ref{fig:01}(a), with a system size of \(L = 6\) and parameters \(\omega_{0}/t = 1\), \(\lambda_{D}=2 / 3\), \(\beta = 6\) and \(\mu=-4\). Fig. \ref{fig:01}(b) demonstrates the influence of different disorder strengths on the charge-correlation function \(c(\boldsymbol{r})\). As the strength of the disorder \(\Delta\) increases, the oscillation amplitude and pattern of the charge-correlation function \(c(\boldsymbol{r})\) change significantly. When \(\Delta = 0\), \(c(\boldsymbol{r})\) exhibits a specific oscillation pattern, indicating that there is a certain correlation rule for charges on the lattice in this case. The critical temperature is \(\beta_c\approx5.8\) in this case \cite{PhysRevLett.122.077602}, so the system enters the CDW phase under the parameter condition of \(\beta = 6\). When \(\Delta\) increases to \(0.5\), the oscillation amplitude of \(c(\boldsymbol{r})\) decreases slightly. As \(\Delta\) increases further to \(1.0\) and \(1.5\), the changes in \(c(\boldsymbol{r})\) become more pronounced, which reflects that the increase in disorder strength disrupts the original charge distribution pattern and interferes with the charge correlations. Essentially, the disorder strength \(\Delta\) represents the degree of interference of impurities or defects in the system on the motion of electrons. The variation of \(c(\boldsymbol{r})\) with \(\Delta\) indicates that disorder affects the localization and delocalization of electrons in the lattice. The stronger disorder makes the electrons more likely to localize, thereby weakening the long-range charge correlations in the lattice.

\begin{figure}[t]
\centering
\includegraphics[width=1.0\columnwidth]{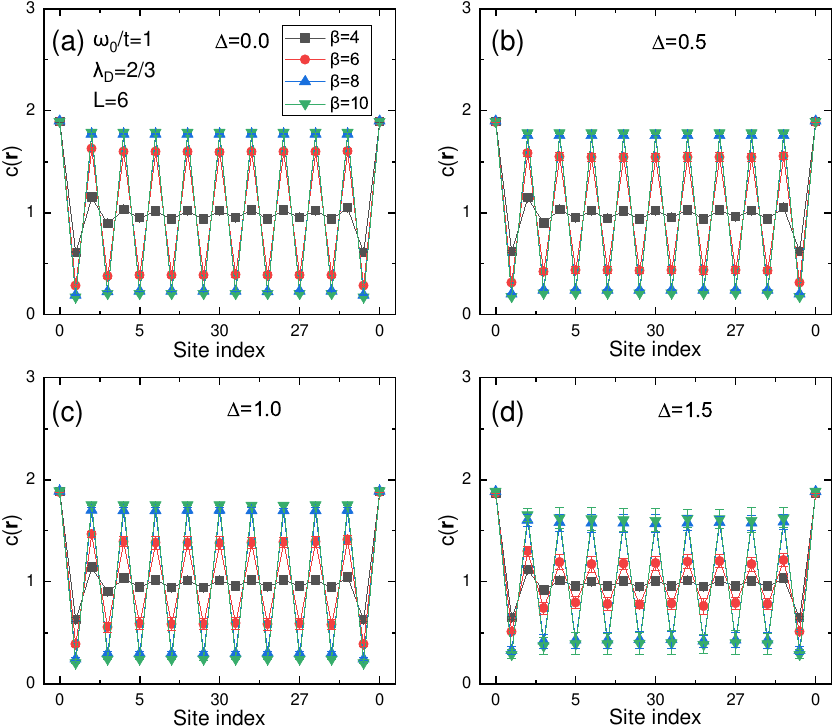}
\caption{The charge-correlation function \(c(\boldsymbol{r})\) along a specific real-space path shown as the red line in FIG.\ref{fig:01}(a). Under the conditions of a system size \(L = 6\) and fixed parameters \(\omega_{0}/t= 1\), \(\lambda_{D}=2 / 3\), with different temperature parameters \(\beta\) and different disorder strengths are considered: (a) \(\Delta = 0.0\), (b) \(\Delta = 0.5\), (c) \(\Delta = 1.0\), (d) \(\Delta = 1.5\).} %
 \label{fig:02}
 \end{figure}

For a more in-depth exploration of disorder effects, in Fig. \ref{fig:02}, we present more detailed results of the charge-correlation function \(c(\boldsymbol{r})\) corresponding to \(\beta = 4\), \(\beta = 6\), \(\beta = 8\), and \(\beta = 10\) under different disorder strengths \(\Delta\). Fig. \ref{fig:02}(a) demonstrates the influence of different temperature parameters \(\beta\)  on the charge-correlation function \(c(\boldsymbol{r})\) when the disorder strength \(\Delta = 0\). In this case, both \(\beta = 8\) and \(\beta = 10\) are at saturation values, corresponding to the CDW phase, \(\beta = 6\) is close to the saturation value, and only \(\beta = 4\) has not entered the CDW phase. At lower temperatures (larger \(\beta\) values), the oscillatory characteristics of the charge-correlation function indicate that the interaction between electrons is enhanced, and the distribution of charges on the lattice tends to be more ordered. Compared with the case of \(\Delta = 0\) in Fig. \ref{fig:02}(a), there is no significant change in \(c(\boldsymbol{r})\) in Fig. \ref{fig:02}(b). In Fig. \ref{fig:02}(c) and Fig. \ref{fig:02}(d), when \(\Delta\) increased to $1.0$ and $1.5$, for \(\beta = 6\), the oscillation amplitude of \(c(\boldsymbol{r})\) decreases significantly; the saturation values of \(c(\boldsymbol{r})\) corresponding to \(\beta = 8\) and \(\beta = 10\) also decrease; and the line for \(\beta = 4\), \(c(\boldsymbol{r})\approx\rho^{2} = 1\) indicates no obvious CDW order. This indicates that as \(\Delta\) increases, \(\beta_{c}\) rises from \(\beta_{c}\approx5.8\) when \(\Delta = 0\) to the region of \(6\leq\beta\leq8\). The introduction of disorder strength subjects electron motion to additional scattering, which partially disrupts the charge-correlation characteristics dominated by temperature.

\subsection{The CDW structure factor}

\begin{figure}[t]
\centering
\includegraphics[width=1.0\columnwidth]{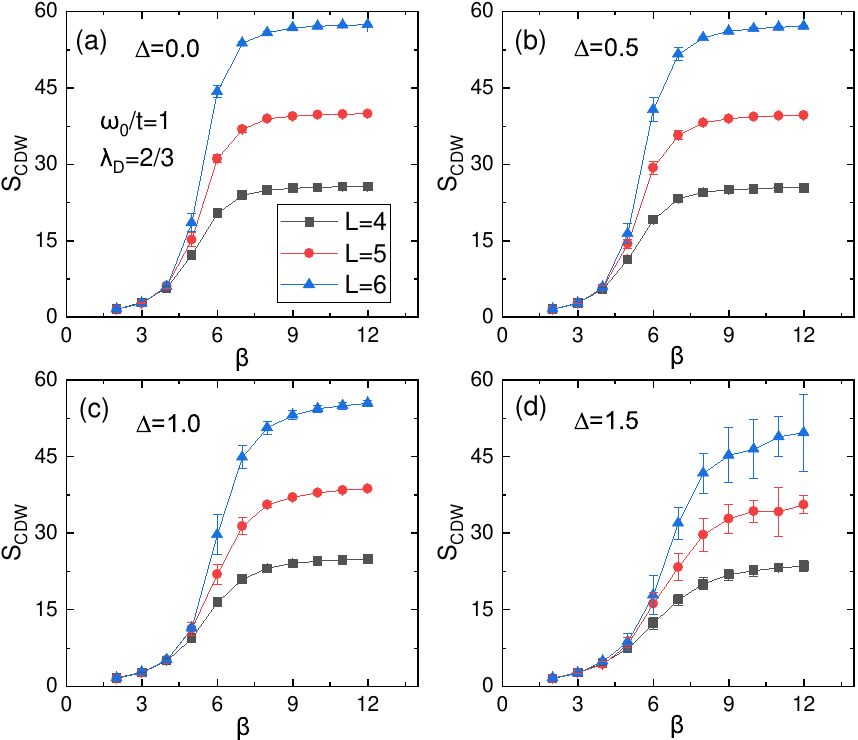}
\caption{ The variation of the CDW structure factor \(S_{CDW}\) with the inverse temperature \(\beta\) for different lattice sizes, with \(\lambda_{D} = 2/3\), \(\omega_{0} = 1\), \(\rho = 1\), and disorder strength \(\Delta\) being (a) \(\Delta = 0\), (b) \(\Delta = 0.5\), (c) \(\Delta = 1.0\), (d) \(\Delta = 1.5\) respectively. } %
 \label{fig:03}
 \end{figure}

To further conduct a quantitative analysis of the strength of CDW, we introduce the CDW structure factor. In Fig. \ref{fig:03}(a), when the disorder strength \(\Delta = 0\), for different lattice sizes, as \(\beta\) increases, \(S_{CDW}\) gradually increases. During the process of rising from the plateau in the region of smaller \(\beta\) to the plateau in the region of larger \(\beta\), there exists a region where \(S_{CDW}\) rises sharply. Curves for different lattice sizes are closer when \(\beta\) is small, and as \(\beta\) increases, the curves gradually separate, indicating that the lattice size also has a significant impact on the variation of the charge structure factor in the low-temperature region. This variation reflects the synergistic effect of temperature and lattice size on the charge order of the system in the absence of disorder. In Fig. \ref{fig:03}(b), when the disorder strength \(\Delta\) is adjusted from 0 to \(0.5\), the overall trend of the curve does not change significantly, indicating that weaker disorder does not significantly affect the charge order. In Fig. \ref{fig:03}(c), the disorder strength is increased to \(1.0\), and the curve shows obvious differences compared with those in Fig. \ref{fig:03}(a) and Fig. \ref{fig:03}(b). The growth rate and saturation value of \(S_{CDW}\) are significantly affected at different \(\beta\), and disorder makes the response of the charge structure factor to temperature more complex. This reflects that disorder with sufficient strength will have a destructive effect on the charge order. In Fig. \ref{fig:03}(d), overly strong disorder leads to a significant change in the variation law of \(S_{CDW}\), and both the saturation value and variation trend of the charge structure factor are strongly disturbed. The maximum value of the error bar also changes from the rapid-rising region of \(S_{CDW}\) in Fig. \ref{fig:03}(b) and Fig. \ref{fig:03}(c) to the region of larger \(\beta\). This indicates the interfering effect of strong disorder on the charge order, so that the influence of temperature and lattice size on the charge structure factor is seriously weakened.

\begin{figure*}[t]
\centering
\includegraphics[width=2.0\columnwidth]{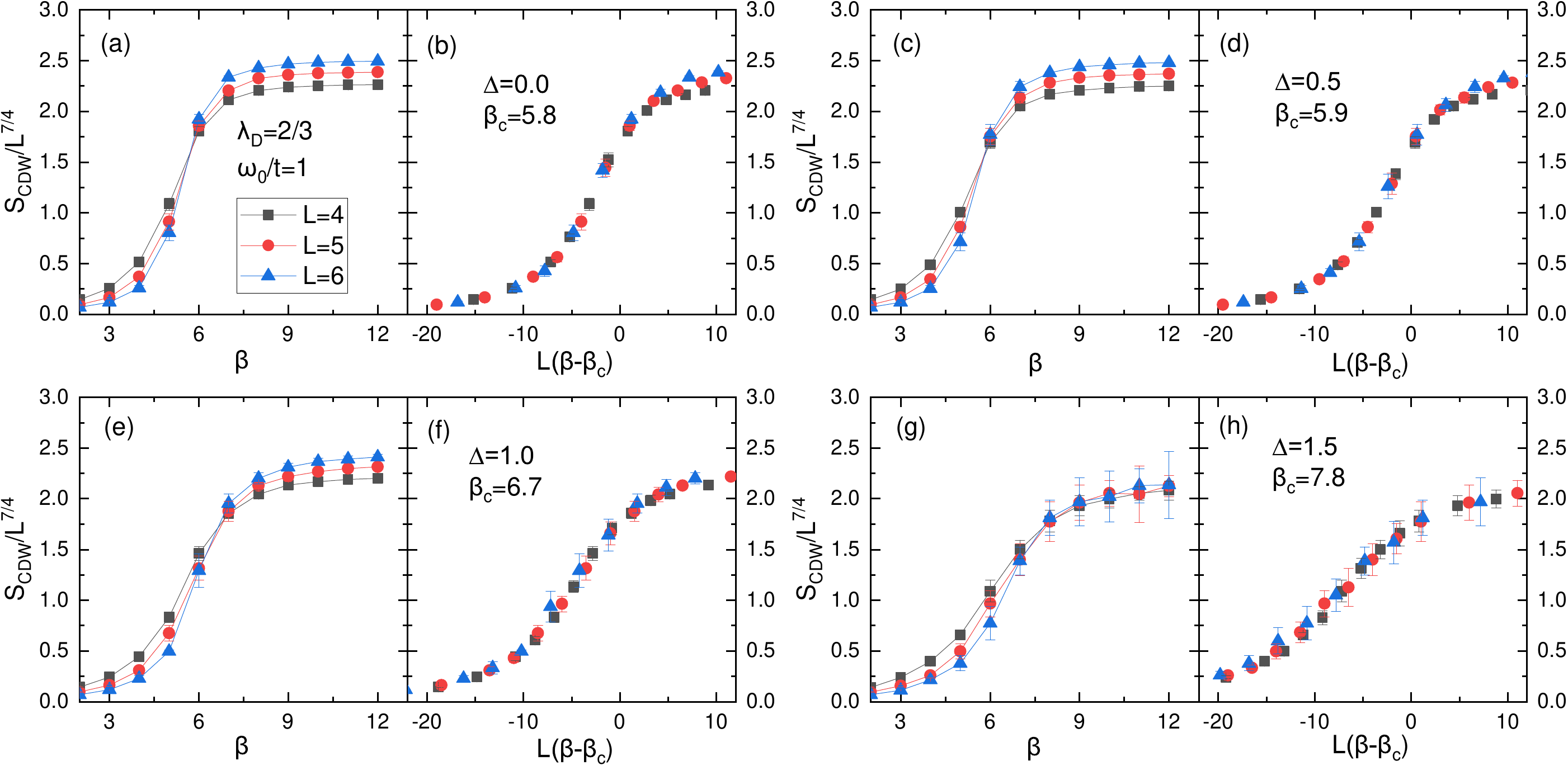}
\caption{ The crossing plots and best data collapse of $S_{\rm CDW}/L^{7/4}$ for different disorder strengths $\Delta$. (a), (c), (e), (g) show the variation of $S_{\rm CDW}/L^{7/4}$ with $\beta$ for lattice sizes $L=4, 5, 6$; (b), (d), (f), (h) present the corresponding data collapse (via 2D Ising critical exponents) as a function of $L(\beta-\beta_c)$. The extracted critical temperatures are as follows: (a), (b) $\beta_c=5.8$ for $\Delta=0.0$, (c), (d) $\beta_c=5.9$ for $\Delta=0.5$, (e), (f) $\beta_c=6.7$ for $\Delta=1.0$, and (g), (h) $\beta_c=7.8$ for $\Delta=1.5$. Here, $\lambda_{\rm D}=2/3$ and $\omega_0/t=1$.}%
 \label{fig:04}
 \end{figure*}

Employing the critical exponents $\gamma = 7/4$ and $\nu = 1$ of the two-dimensional Ising model, a finite-size scaling analysis of the CDW structure factor enables precise determination of the critical temperature $\beta_c$ \cite{Binder1981}, as illustrated in Fig.~\ref{fig:04}. For the clean system ($\Delta = 0.0$), Fig.~\ref{fig:04}(a) presents $S_{\mathrm{CDW}}/L^{7/4}$ as a function of $\beta$ for lattice sizes $L = 4, 5, 6$, while Fig.~\ref{fig:04}(b) shows the corresponding finite-size scaling collapse: data for all $L$ converge cleanly around $\beta_c = 5.8$, confirming the reliability of this critical temperature. This $\beta_c = 5.8$ serves as a benchmark for analyzing modifications of critical behavior induced by disorder. When weak disorder ($\Delta = 0.5$) is introduced, Fig.~\ref{fig:04}(c) depicts $S_{\mathrm{CDW}}/L^{7/4}$ versus $\beta$, and Fig.~\ref{fig:04}(d) displays its scaling collapse. The overall convergence trend resembles that of the clean system, but $\beta_c$ shifts slightly to 5.9, and critical fluctuations are enhanced (manifested as a broader spread of the curves)---a result consistent with weak disorder perturbing, yet not disrupting, charge-order correlations. For moderate disorder ($\Delta = 1.0$), Fig.~\ref{fig:04}(e) illustrates $S_{\mathrm{CDW}}/L^{7/4}$ as a function of $\beta$, and Fig.~\ref{fig:04}(f) presents its scaling collapse. Here, the data converge around $\beta_c = 6.7$, though the collapse is less tight than for weaker disorder. This phenomenon originates from disorder-induced electron localization (with localization length $\xi \approx L$): scattering events truncate long-range charge order, yet conventional scaling analysis using 2D Ising exponents still yields a consistent critical temperature. Under strong disorder ($\Delta = 1.5$), Fig.~\ref{fig:04}(g) shows $S_{\mathrm{CDW}}/L^{7/4}$ versus $\beta$, and Fig.~\ref{fig:04}(h) presents its scaling behavior. Although a value of $\beta_c = 7.8$ is extracted, the data collapse deteriorates markedly: curves for different $L$ diverge across the entire $\beta$ range, and error bars expand substantially (especially for $\beta > 8$). This indicates that strong disorder completely disrupts long-range charge-order correlations, trapping the system in a ``disorder-frozen'' state---where the effects of temperature and lattice size are overwhelmed by strong scattering potentials. Consequently, conventional finite-size scaling becomes only partially applicable, and the enlarged error bars further corroborate the non-equilibrium fluctuation characteristics of charge order under strong disorder.

\subsection{Phase diagram}

\begin{figure}[t]
  \centering
  \includegraphics[width=1.0\columnwidth]{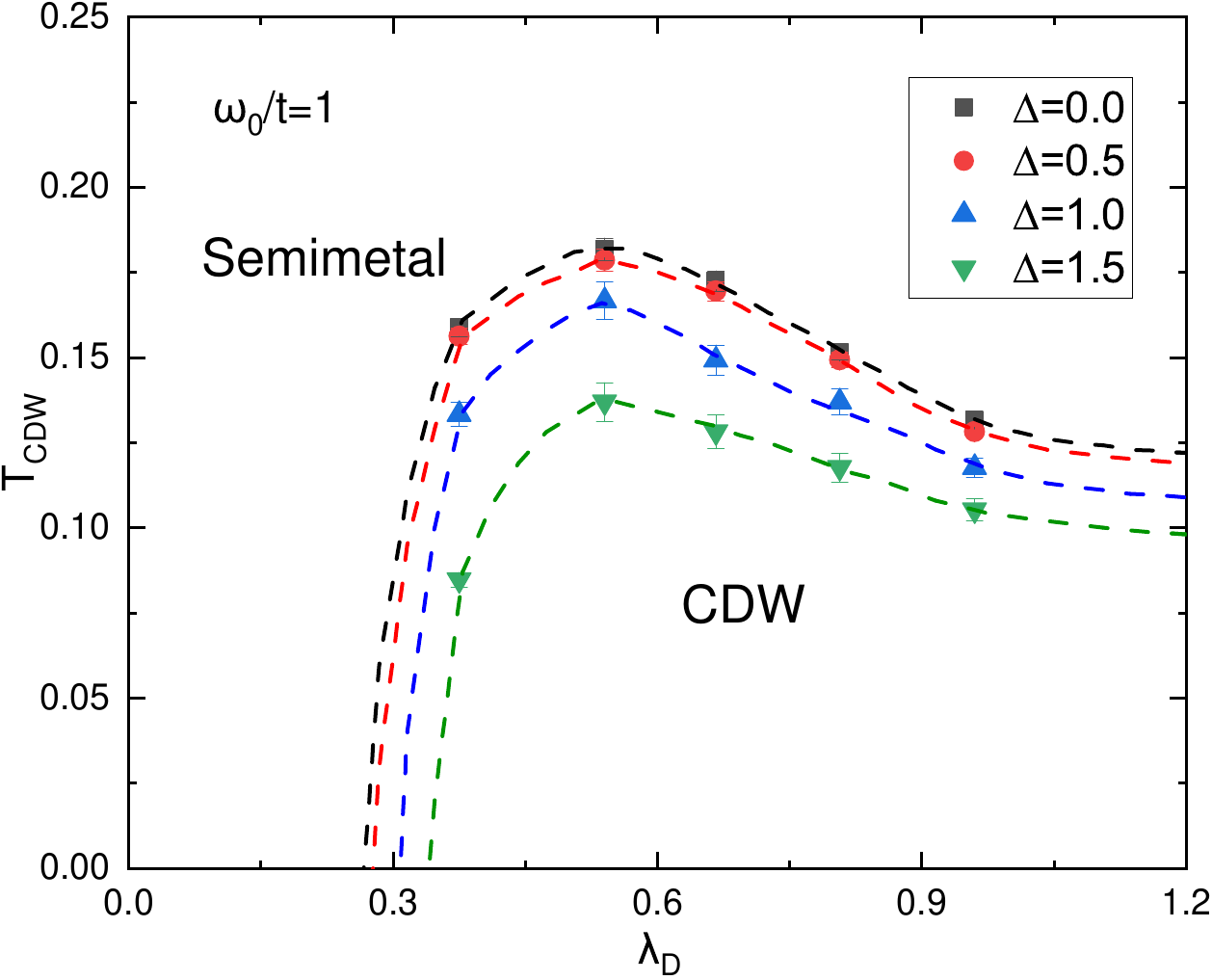}
  \caption{Critical temperature $T_c = 1/\beta_c$ for the CDW phase as a function of the dimensionless electron-phonon coupling strength $\lambda_D$. Results are shown for several fixed values of the disorder strength $\Delta$. Parameters are $\omega_0/t = 1$ and half-filling $\rho = 1$.}
  \label{fig:cdw_phase_diagram}
\end{figure}

To fully characterize the influence of disorder on the stability of CDW order, we construct a phase diagram, as depicted schematically in Fig.~\ref{fig:cdw_phase_diagram}, based on the finite-size scaling analyses performed in this work (Fig.~\ref{fig:04} for \(\lambda_D = 2/3\) and Appendix~\ref{app:appendix_a} for additional coupling strengths). The diagram plots the critical temperature \(T_c = 1/\beta_c\) against the dimensionless electron-phonon coupling strength \(\lambda_D\), with separate curves representing different fixed disorder strengths \(\Delta = 0.0\), \(0.5\), \(1.0\), and \(1.5\). The other parameters are fixed as: phonon frequency \(\omega_0/t = 1\), and half-filling \(\rho = 1\).

In the clean limit, \(T_c\) exhibits a non-monotonic dependence on \(\lambda_D\), similar to findings in prior studies of the Holstein model. As \(\lambda_D\) increases from the critical region, \(T_c\) rises rapidly, reaches a maximum at an intermediate coupling strength (approximately \(\lambda_D \sim 0.4-0.5\) for the honeycomb lattice), and then gradually decreases for larger \(\lambda_D\). This dome-shaped behavior originates from the interplay between two opposing effects of the electron-phonon interaction: the enhancement of CDW coherence and the mass renormalization of electrons, which reduces their kinetic energy and the efficacy of virtual hopping processes underlying CDW formation.

Introducing disorder modifies this profile. Although the overall dome shape persists for weak to moderate disorder (\(\Delta = 0.5, 1.0\)), the curves are progressively shifted downward. Furthermore, the suppression of \(T_c\) by disorder is more pronounced in the strong-coupling regime (larger \(\lambda_D\)) compared to the intermediate-coupling regime near the maximum of \(T_c\). This suggests that disorder and the electron-phonon coupling act as competing mechanisms in determining the charge order stability, with disorder exerting a dominant destabilizing effect when the electron-phonon interaction is very strong.

For strong disorder, the \(T_c\) curve is significantly flattened and lowered. The data collapse in the finite-size scaling analysis also deteriorates at this disorder strength (see Fig.~\ref{fig:04}(h) and Appendix~\ref{app:appendix_a}), indicating that the long-range CDW correlations are severely disrupted. The system approaches a disorder-dominated, ``frozen" state where the phase transition boundary becomes broad and the conventional scaling analysis is only partially applicable.

The phase diagram also bears implications for the quantum critical point (QCP). In the clean system, a QCP is expected at a finite critical coupling \(\lambda_{c}\) (estimated to be \(\lambda_{D} \approx 0.28\) for the honeycomb Holstein model), below which long-range CDW order is absent at zero temperature. Our results indicate that disorder pushes the effective critical coupling to higher values. That is, for a fixed \(\lambda_D > \lambda_{c}\), increasing disorder can drive the system from an ordered CDW phase into a disordered semimetal, effectively expanding the paramagnetic (semimetal) region of the phase diagram. Under strong disorder, the sharp QCP may be smeared into a quantum critical region, characterized by enhanced fluctuations and the failure of clean-lattice critical scaling.

\subsection{Electron kinetic energy, DC conductivity and spectral function}

\begin{figure}[t]
\centering
\includegraphics[width=1.0\columnwidth]{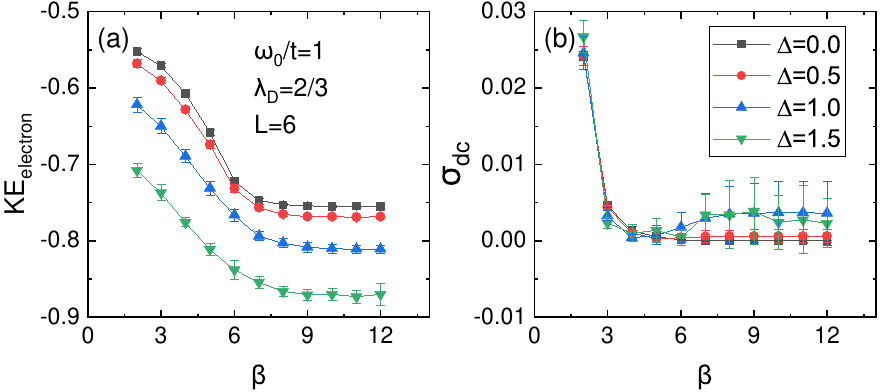}
\caption{(a) Electron kinetic energy and (b) dc conductivity as functions of the inverse temperature, and for different disorder strengths, at fixed \(L = 6\), \(\omega_{0} = 1\), and \(\lambda_{D} = 2/3\). } %
 \label{fig:05}
 \end{figure}

To further understand the relationship between the crossover phenomenon of \(S_{CDW}/L^{\gamma/\nu}\) and the kinetic energy of electrons, we examined how the kinetic energy of electrons varies with temperature under different disorder strengths \cite{PhysRevB.57.6376,PhysRevB.103.L060501}. In  Fig. \ref{fig:05}(a), when \(\Delta = 0\), the kinetic energy of electrons monotonically decreases as \(\beta\) increases, reflecting the formation of a CDW phase at low temperatures, electrons become localized due to lattice distortion, and their kinetic energy is suppressed. The CDW phase reduces electron mobility through charge ordering. Scattering introduced by weak disorder slightly reduces the kinetic energy of electrons. When the disorder strength increases above 1.0, the slope of the kinetic energy as a function of \(\beta\) decreases significantly, indicating that strong disorder inhibits CDW formation by localizing electrons, preventing electrons from reducing their kinetic energy through virtual hopping. The change in kinetic energy reflects the competition between disorder and electron-phonon coupling. Under weak disorder, CDW still dominates the evolution of kinetic energy, while moderate to strong disorder disrupts charge correlations through Anderson localization, making the kinetic energy more determined by the disorder potential rather than the temperature-driven ordering process.  Fig. \ref{fig:05}(b) shows that as the temperature decreases, the dc conductivity decreases rapidly, with \(d\sigma_{dc}/dT > 0\), indicating insulating behavior for all values of \(\Delta\) \cite{PhysRevLett.83.4610,PhysRevB.54.R3756,PhysRevB.103.L060501}.
We note that the non-monotonic behavior of \(\sigma_{\rm dc}\) as a function of \(\beta\) (observed at large disorder \(\Delta \geq 1.0\)) arises from large statistical fluctuations: strong disorder enhances quantum fluctuations, slowing the convergence of DQMC simulations and increasing sampling error, evidenced by widened error bars in the corresponding data. This non-monotonic feature has no physical significance and is purely a sampling artifact.

\begin{figure}[t]
\centering
\includegraphics[width=1.0\columnwidth]{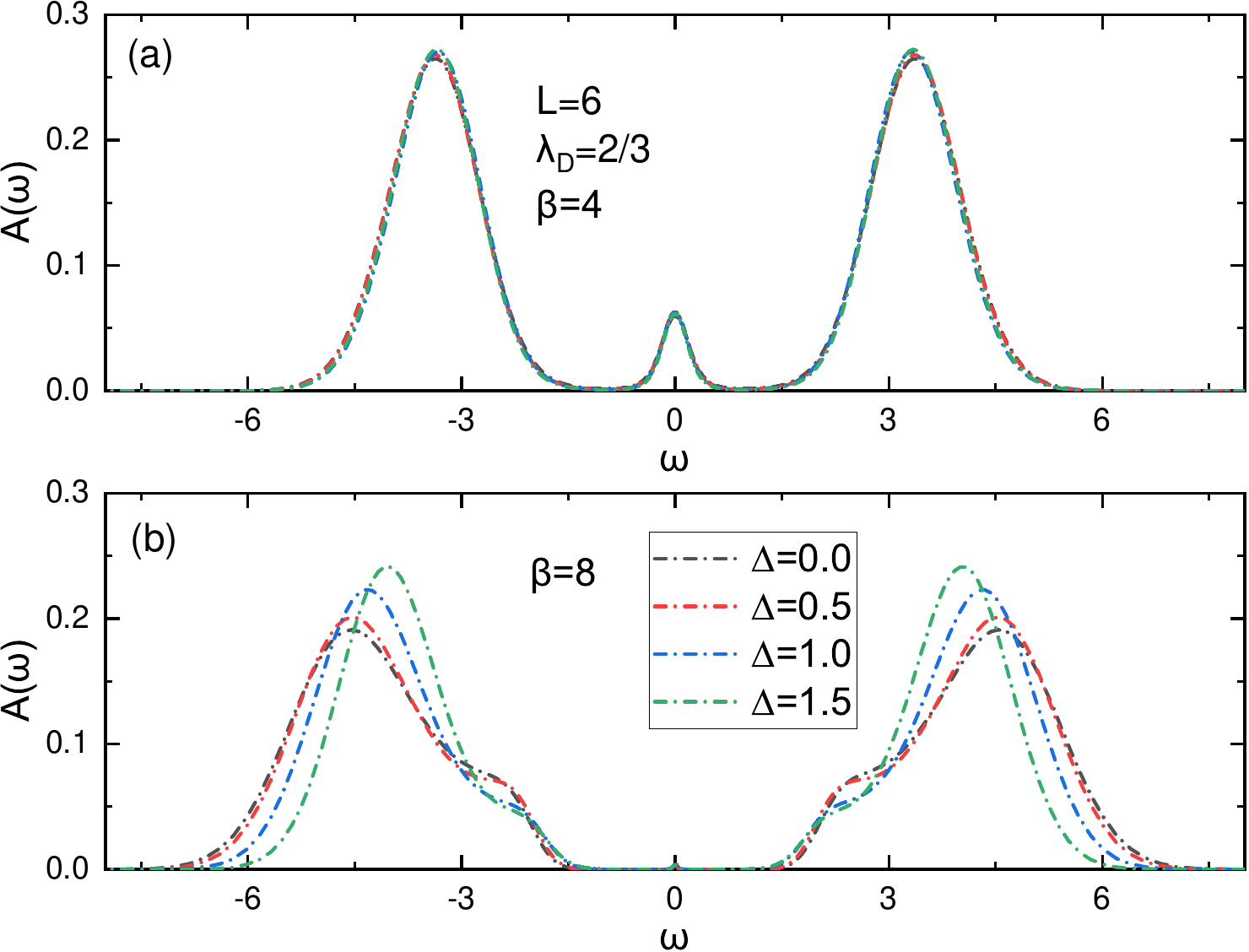}
\caption{Spectral function $A(\omega)$ for different disorder strengths $\Delta$ at fixed lattice size $L=6$, dimensionless electron-phonon coupling $\lambda_D=2/3$, phonon frequency ratio $\omega_0/t=1$, and half-filling $\rho=1$. (a) Results at $\beta=4$ (higher temperature), and (b) results at $\beta=8$ (lower temperature).} %
 \label{fig:06}
 \end{figure}

To reveal the microscopic regulatory mechanism of disorder on CDW order, we calculated the spectral function $A(\omega)$, with the results shown in Fig.~\ref{fig:06}. The system parameters were fixed as follows: lattice size $L=6$, dimensionless electron-phonon coupling $\lambda_D=2/3$, phonon frequency ratio $\omega_0/t=1$, and half-filling $\rho=1$. Fig.~\ref{fig:06}(a) corresponds to a higher temperature, where thermal fluctuations dominate the electronic behavior. The $A(\omega)$ curves exhibit broadened and dispersed features with no significant differences across different disorder strengths. This indicates that at high temperatures, the effect of thermal fluctuations far outweighs that of disorder, and the modulating effect of disorder on the spectral function is masked. Fig.~\ref{fig:06}(b) corresponds to a lower temperature, where the role of disorder becomes prominent as thermal fluctuations weaken: a strict CDW gap exists in the clean system; as $\Delta$ increases to 0.5, 1.0, and 1.5, the gap is progressively broadened and suppressed. This significant broadening of the gap edge and the suppression of the coherence peaks confirm that disorder disrupts the long-range CDW order, consistent with the mechanism of Anderson localization. These results are consistent with the conclusions from the CDW structure factor and electron kinetic energy: at high temperatures, the effect of disorder is masked, while at low temperatures, the competition between electron-phonon coupling and disorder dominates the gap's degradation, providing microscopic support for the suppression of CDW order by disorder.

\section{Summary}
This study investigates the effect of disorder on the CDW order in the Holstein model on the honeycomb lattice using DQMC simulations. Disorder is introduced via random hopping amplitudes $t_{ij}$, which are uniformly distributed around a mean value. The primary finding is that disorder suppresses the CDW phase, as evidenced by a reduction in the CDW structure factor $S_{\text{CDW}}$ and an increase in the critical inverse temperature $\beta_c$, indicating that stronger disorder destabilizes the CDW phase and renders it more susceptible to thermal fluctuations. The analysis of charge-density correlations and finite-size scaling reveals that moderate to strong disorder disrupts long-range CDW order through Anderson localization, leading to the failure of conventional critical scaling. Additionally, disorder suppresses electron kinetic energy and dc conductivity, further impairing electronic transport. Spectral function calculations show that disorder broadens and eventually closes the CDW energy gap. These results provide important insights into the stability and critical behavior of correlated electronic phases in disordered 2D Dirac systems.

\begin{acknowledgments}G.C.L and L.F.Z were supported by the Fundamental Research Funds for the Central Universities from the Beijing University of Posts and Telecommunications under Grant No. 2024ZCJH13 and 2025JCTP08. This work was supported by NSFC (12474218) and Beijing Natural Science Foundation (No. 1242022 and 1252022).  We also supported by The Key Laboratory of Multiscale Spin Physics with Contract No. SPIN2024N04. We acknowledge computational support from the HSCC of Beijing University of Posts and Telecommunications.
\end{acknowledgments}

\appendix

\section{FINITE-SIZE SCALING ANALYSES FOR DIFFERENT ELECTRON-PHONON COUPLING STRENGTHS}
\label{app:appendix_a}

\begin{figure}[t]
  \centering
  \includegraphics[width=1.0\columnwidth]{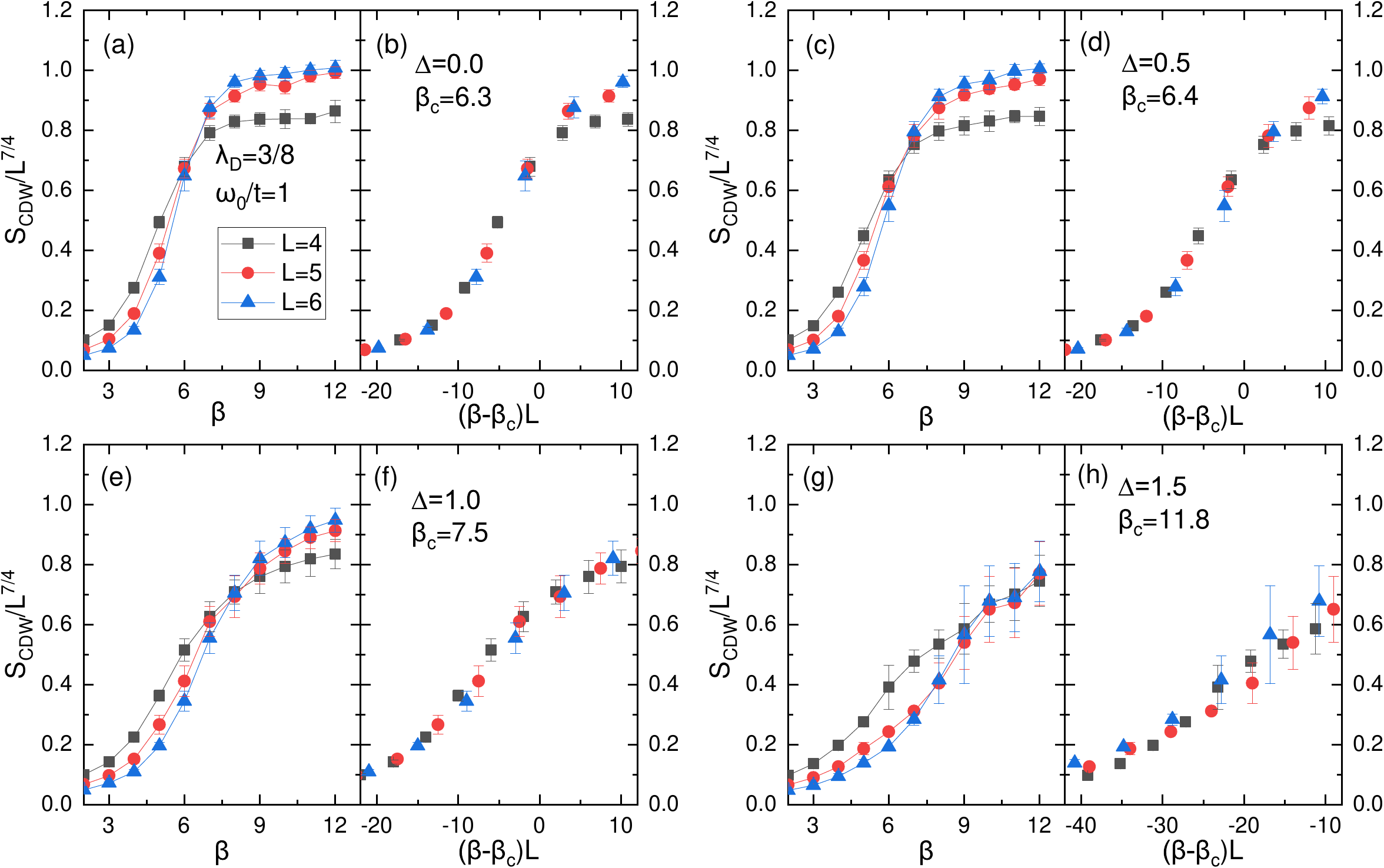}
  \caption{The crossing plots and best data collapse of $S_{\rm CDW}/L^{7/4}$ for different disorder strengths $\Delta$. (a), (c), (e), (g) show the variation of $S_{\rm CDW}/L^{7/4}$ with $\beta$ for lattice sizes $L=4, 5, 6$; (b), (d), (f), (h) present the corresponding data collapse (via 2D Ising critical exponents) as a function of $L(\beta-\beta_c)$. Here, $\lambda_{\rm D}=3/8$ and $\omega_0/t=1$.}
  \label{fig:appendix_lambda1_5}
\end{figure}

\begin{figure}[t]
  \centering
  \includegraphics[width=1.0\columnwidth]{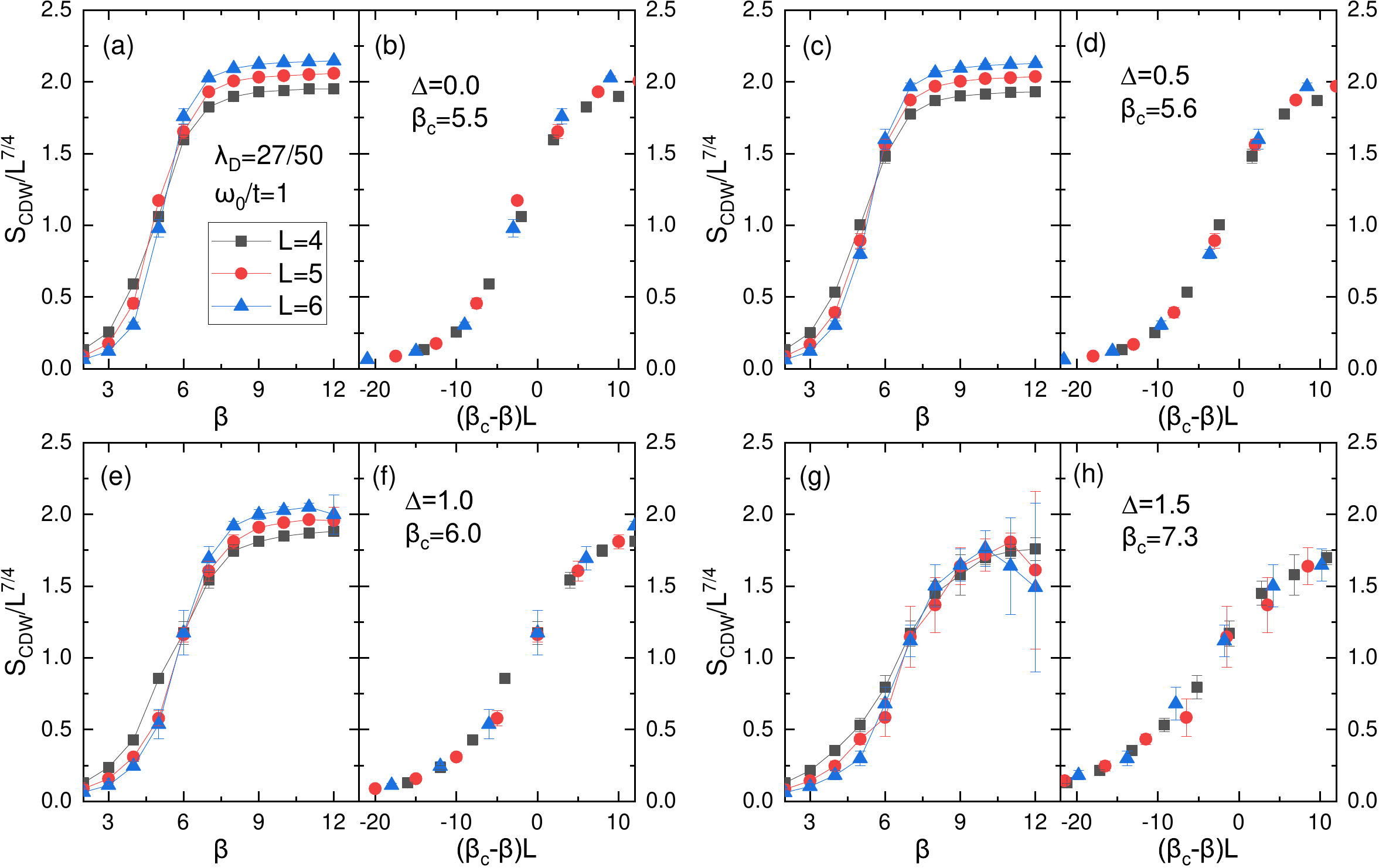}
  \caption{The crossing plots and best data collapse of $S_{\rm CDW}/L^{7/4}$ for different disorder strengths $\Delta$. Here, $\lambda_{\rm D}=27/50$ and $\omega_0/t=1$.}
  \label{fig:appendix_lambda1_8}
\end{figure}

\begin{figure}[t]
  \centering
  \includegraphics[width=1.0\columnwidth]{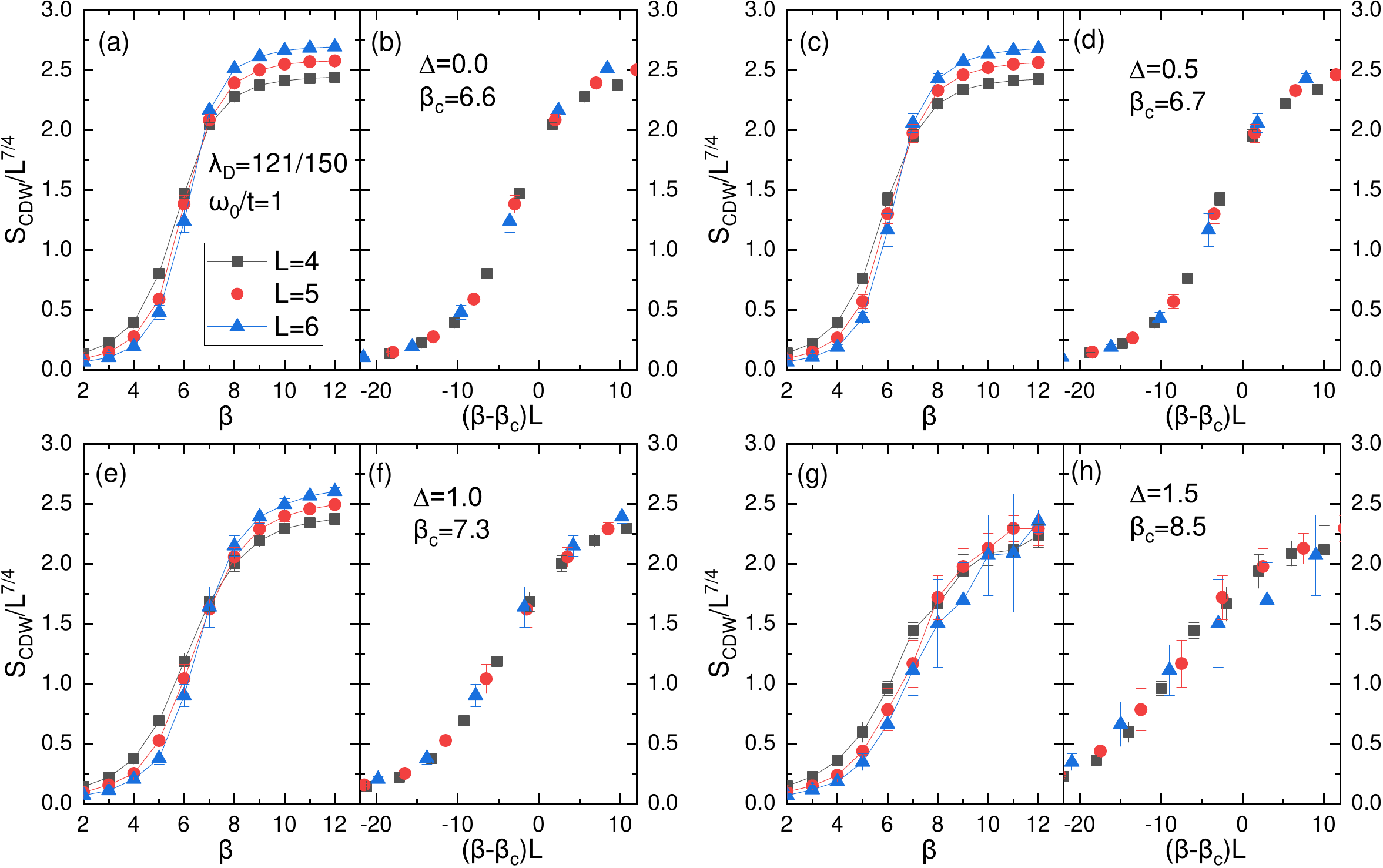}
  \caption{The crossing plots and best data collapse of $S_{\rm CDW}/L^{7/4}$ for different disorder strengths $\Delta$. Here, $\lambda_{\rm D}=24/25$ and $\omega_0/t=1$.}
  \label{fig:appendix_lambda2_2}
\end{figure}

\begin{figure}[t]
  \centering
  \includegraphics[width=1.0\columnwidth]{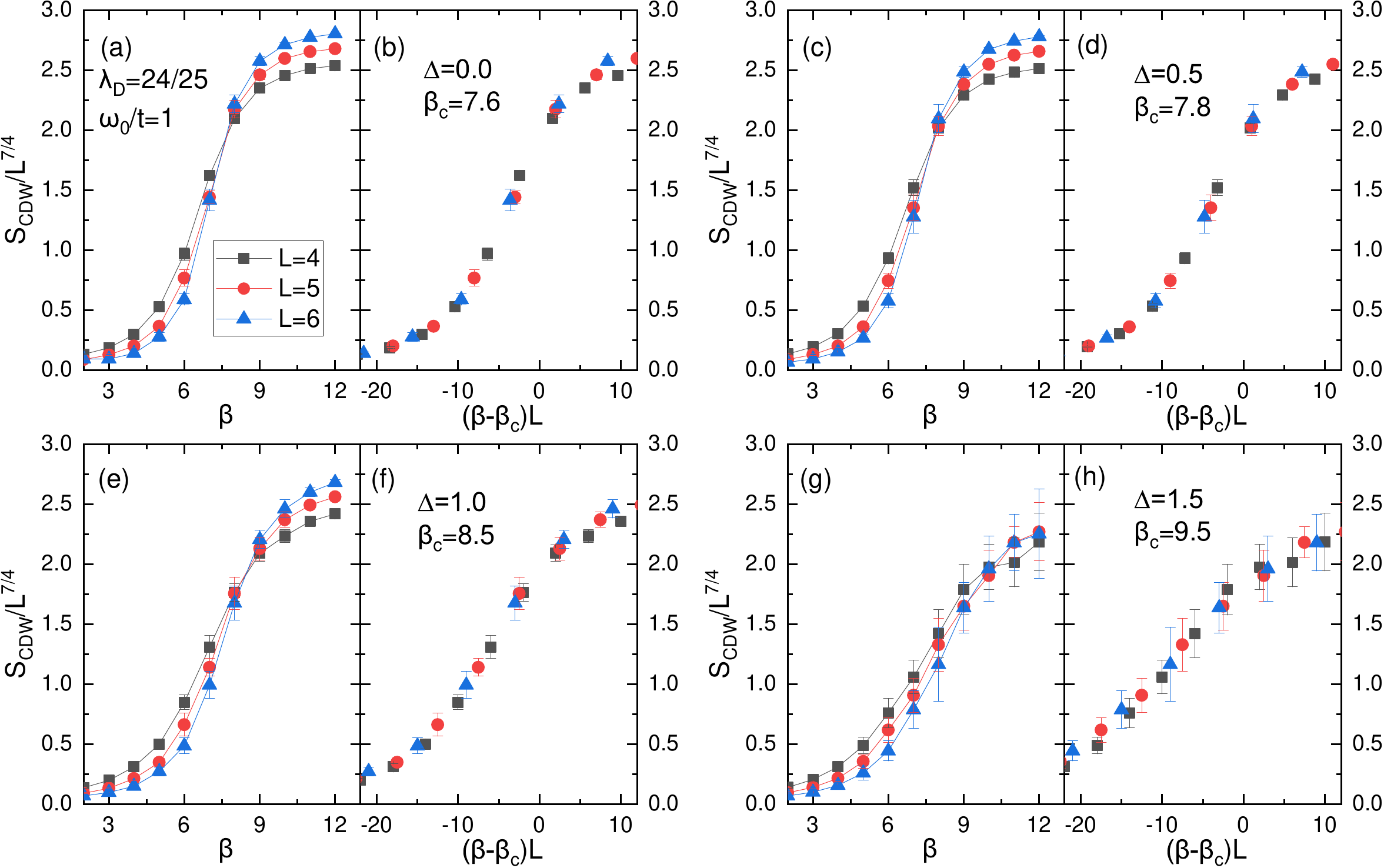}
  \caption{The crossing plots and best data collapse of $S_{\rm CDW}/L^{7/4}$ for different disorder strengths $\Delta$. Here, $\lambda_{\rm D}=121/150$ and $\omega_0/t=1$.}
  \label{fig:appendix_lambda2_4}
\end{figure}

This appendix (including Fig.~\ref{fig:appendix_lambda1_5}, Fig.~\ref{fig:appendix_lambda1_8}, Fig.~\ref{fig:appendix_lambda2_2} and Fig.~\ref{fig:appendix_lambda2_4}) presents supplementary finite-size scaling results for the CDW structure factor $S_{\mathrm{CDW}}$ in systems with increased electron-phonon coupling. We examine four cases corresponding to the original coupling parameters $\lambda = 1.5,\; 1.8,\; 2.2,\; 2.4$, which translate to the following dimensionless coupling strengths: $\lambda_D = \lambda^2/(\omega_0^2 W) = 3/8,\; 27/50,\; 121/150,\; 24/25$, respectively. All analyses employ the critical exponents $\gamma = 7/4$ and $\nu = 1$ of the two-dimensional Ising model \cite{Binder1981}, consistent with the main text. The fixed parameters throughout are $\omega_0/t = 1$, half-filling $\rho = 1$, lattice sizes $L = 4, 5, 6$, and disorder strengths $\Delta = 0.0,\; 0.5,\; 1.0,\; 1.5$. These results further clarify the combined influence of EPC and disorder on the CDW critical behavior, extending the analysis presented in the main text for $\lambda_D = 2/3$.

\section{\textit{S}-WAVE PAIRING SUSCEPTIBILITY ANALYSIS}
\label{app:appendix_b}

\begin{figure}[t]
  \centering
  \includegraphics[width=0.9\columnwidth]{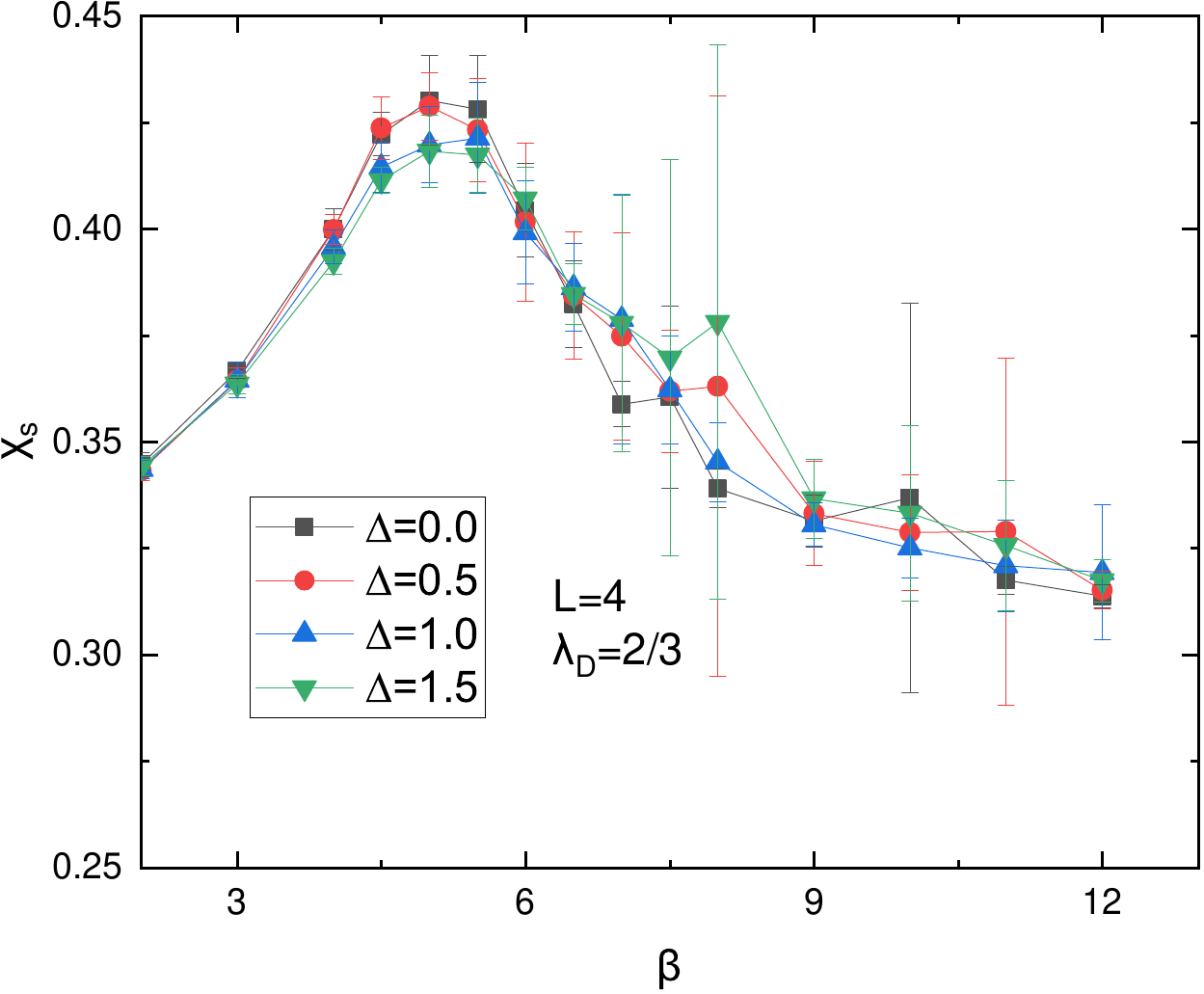}
  \caption{The \textit{s}-wave pairing susceptibility \(\chi_s\) as a function of inverse temperature \(\beta\) for \(L = 4\), \(\lambda_D = 2/3\), and disorder strengths \(\Delta = 0.0,\; 0.5,\; 1.0,\; 1.5\).}
  \label{fig:swave_susceptibility}
\end{figure}

To investigate superconducting pairing correlations in the disordered honeycomb Holstein model, we compute the \textit{s}-wave pairing susceptibility \(\chi_s\), a key quantity for characterizing \textit{s}-wave superconducting correlations in two-dimensional systems. It is defined as
\[
\chi_{s} = \frac{1}{N} \int_{0}^{\beta} d\tau\,
\bigl\langle \Delta(\tau) \Delta^\dagger(0) \bigr\rangle,
\]
where \(\Delta(\tau) = \sum_i \hat{d}_{i\uparrow}(\tau) \hat{d}_{i\downarrow}(\tau)\) is the imaginary-time dependent \textit{s}-wave pairing operator, and \(\hat{d}_{i\sigma}(\tau) = e^{\mathcal{H}\tau} \hat{d}_{i\sigma} e^{-\mathcal{H}\tau}\) denotes the imaginary-time evolved electron annihilation operator at site \(i\) with spin \(\sigma\).

All calculations use the same core parameters as the main text: lattice size \(L = 4\), dimensionless electron–phonon coupling \(\lambda_D = 2/3\), phonon frequency ratio \(\omega_0/t = 1\), and half-filling \(\rho = 1\). Disorder strengths \(\Delta = 0.0,\; 0.5,\; 1.0,\; 1.5\) are varied to examine the influence of disorder on \textit{s}-wave pairing correlations.

Fig.~\ref{fig:swave_susceptibility} displays the \textit{s}-wave pairing susceptibility $\chi_s$ as a function of $\beta$ for different disorder strengths $\Delta$. The key result is that the $\chi_s$ curves for $\Delta = 0.0$, $0.5$, $1.0$, and $1.5$ nearly coincide across the entire temperature range. This overlap indicates that the temperature dependence of $\chi_s$ is virtually identical regardless of the disorder strength. Most importantly, even at the lowest temperatures ($\beta \gtrsim 9$), there is no measurable difference between the susceptibility of the clean system and that of the strongly disordered system ($\Delta = 1.5$).

This result clearly demonstrates that disorder has no significant impact on the \textit{s}-wave pairing susceptibility in this system. This finding complements the main-text conclusion that disorder strongly suppresses CDW correlations, highlighting the selective role of disorder: it disrupts charge order while leaving \textit{s}-wave pairing correlations essentially unaffected in the honeycomb Holstein model.

\bibliography{reference}

\end{document}